\documentclass[aps,pra,reprint,onecolumn,notitlepage,showkeys,showpacs,amsfonts,
              amsmath]{revtex4-1}
\usepackage{amsthm}

\theoremstyle{definition}

\newcommand{\secref}[1]{Section~\ref{#1}}

\newcommand{\Secref}[1]{Section~\ref{#1}}

\newcommand{\dd}{\mathrm{d}}

\newcommand{\nbr}[1]{$#1$\nobreakdash-\hspace{0pt}}
\renewcommand{\vec}[1]{\boldsymbol{\mathbf{#1}}}
\providecommand{\abs}[1]{\lvert#1\rvert}

\DeclareMathOperator{\diag}{diag}

\numberwithin{equation}{section}

\begin{document}

\title{Flat minimal quantizations of St\"ackel systems and quantum separability}
\author{Maciej B{\l}aszak}
\email[Electronic address: ]{blaszakm@amu.edu.pl}
\author{Ziemowit Doma{\'n}ski}
\email[Electronic address: ]{ziemowit@amu.edu.pl}
\affiliation{Faculty of Physics, Adam Mickiewicz University\\
             Umultowska 85, 61-614 Pozna{\'n}, Poland}
\author{Burcu Silindir}
\email[Electronic address: ]{burcu.yantir@ieu.edu.tr}
\affiliation{Department of Mathematics, \'{Y}zmir University of Economics\\
             35330, Bal\c{c}ova,\'{Y}zmir, Turkey}
\date{\today}

\begin{abstract}
In this paper, we consider the problem of quantization of classical St\"ackel
systems and the problem of separability of related quantum Hamiltonians. First,
using the concept of St\"ackel transform, all considered systems are expressed
by flat coordinates of related Euclidean configuration space. Then, the
so-called flat minimal quantization procedure is applied in order to construct
an appropriate Hermitian operator in the  respective Hilbert space. Finally, we
distinguish a class of St\"ackel systems which remain separable after any of
admissible flat minimal quantizations.
\end{abstract}



\maketitle

%%%%%%%%%%%%%%%%%%%%%%%%%%%%%%%%%%%%%%%%%%%%%%%%%%%%%%%%%%%%%%%%%%%%%%%%%%%%%%%%
\section{Introduction}
\label{sec:1}
There exists a connection between classical Hamiltonian systems and quantum
systems, through an appropriate quantization procedure \cite{Moyal:1949,%
Bayen:1978a,Bayen:1978b,Dito.Sternheimer:2002}. It is of great
interest to investigate this connection as it could help to
transfer results from classical theory to quantum theory. One of
the particularly interesting problems, is a relation
between integrability and in particular separability  of classical
and quantum systems. Some partial results on that subject can
be found in literature
\cite{Liu:1990,Toth:1995,Harnad:1995,Mykytiuk:1994}. In this paper
we are going to investigate systematically a separability of
quantum systems received from classical St\"ackel systems, i.e.
these systems for which all constants of motion are quadratic in
momenta, by means of an appropriate quantizations. It should be
noted that there is a variety of quantization procedures leading
to different quantum systems \cite{Blaszak:2013b}. In this paper
we are going to focus on a so-called minimal quantization.

In our approach quantization depends on a metric tensor from a
configuration space. With every classical St\"ackel system is
associated a natural metric tensor, which can be used to quantize
such a system. In \cite{Blaszak:2013c} it was shown that so called
Benenti class of St\"ackel systems after minimal quantization
leads to quantum separable systems (the respective system of
stationary Schr\"odinger equations is separable
\cite{Benenti:2002a,Benenti:2002b}). In this paper we are going to
consider the whole family of admissible quantizations of St\"ackel
systems and investigate the problem of their quantum separability.

It is known that for each pair of classical St\"ackel systems
there exists a St\"ackel transform relating them
\cite{Sergyeyev:2008,Blaszak.Marciniak:2012}. Using this fact we
can relate any St\"ackel system with a chosen flat system and
introduce quantization by means of a natural flat metric induced
by that system.

In \secref{sec:2} we refer basic notions about St\"ackel systems
and St\"ackel transform. \Secref{sec:3} contains a description of
minimal quantization procedure. In \secref{sec:4} we investigate a
family of flat minimal quantizations of Benenti class of St\"ackel
systems. In particular, we prove that for any Benenti system,
there exists an \nbr{n}parameter family of minimal flat
quantizations, which preserves quantum separability. In
\secref{sec:5} we investigate flat minimal quantizations of
arbitrary classical St\"ackel system. We receive the result that
all admissible flat minimal quantizations of any non-Benenti class
destroy a quantum separability. \Secref{sec:6} presents a
procedure of deformation of St\"ackel systems so as to preserve
the separability of deformed operators which however destroy their
Hermicity. Finally, in \secref{sec:7}, we illustrate the theory by
few examples.

%%%%%%%%%%%%%%%%%%%%%%%%%%%%%%%%%%%%%%%%%%%%%%%%%%%%%%%%%%%%%%%%%%%%%%%%%%%%%%%%
\section{St\"ackel systems in flat coordinates}
\label{sec:2}
Let us recall basic notions from the theory of separable Hamiltonian systems.
Consider a Liouville-integrable system on a \nbr{2n}dimensional phase space
$(M,\mathcal{P})$, where $\mathcal{P}$ is a non-degenerated Poisson tensor.
Then, there exist $n$ functions $H_i$ in involution with respect to a Poisson
bracket:
\begin{equation}
\{H_i,H_j\}_\mathcal{P}:=\mathcal{P}(\dd H_i,\dd H_j) = 0, \quad
i,j = 1,2,\dotsc,n.
\label{eq:43}
\end{equation}
The functions $H_i$ generate $n$ Hamiltonian dynamic systems
\begin{equation}
u_{t_i} = \mathcal{P}\dd{H_i}, \quad i = 1,2,\dotsc,n, \quad u\in M.
\label{eq:44}
\end{equation}
One of the methods of solving the system of equations \eqref{eq:44} is a
Hamilton-Jacobi method. In this method one linearizes equations \eqref{eq:44}
by performing an appropriate canonical transformation of coordinates
$(q,p) \mapsto (b,a)$, $a_i = H_i$. The generating function $W(q,a)$ of such
canonical transformation is then calculated by solving the Hamilton-Jacobi
equations
\begin{equation}
H_i\left(q_1,\dotsc,q_n,\frac{\partial W}{\partial q_1},\dotsc,
    \frac{\partial W}{\partial q_n}\right) = a_i, \quad i = 1,2,\dotsc,n.
\label{eq:45}
\end{equation}
A system of equations \eqref{eq:45} can be solved by separation of variables,
i.e. we have to find a canonical transformation $(q,p) \mapsto (\lambda,\mu)$
to a new coordinate system $(\lambda,\mu)$, called separation coordinates, in
which \eqref{eq:45} separates to a system of $n$ decoupled ordinary differential
equations, which in turn can be solved by quadratures. In other words, in
separation coordinates $(\lambda,\mu)$ there exist the following relations
\begin{gather}
\varphi_i(\lambda_i,\mu_i;a_1,\dotsc,a_n) = 0, \quad i = 1,2,\dotsc,n
\nonumber \\
a_i \in \mathbb{R}, \quad
\det\left[\frac{\partial \varphi_i}{\partial a_j}\right] \neq 0,
\label{eq:46}
\end{gather}
such that each of these relations involves only a single pair of canonical
coordinates. The relations \eqref{eq:46} are called separation relations
\cite{Sklyanin:1995, Blaszak:2009}. In this paper we consider
Liouville-integrable systems having separation relations in the following form
\begin{equation}
H_1 \lambda_i^{\gamma_1} + H_2 \lambda_i^{\gamma_2} + \dotsb
+ H_n \lambda_i^{\gamma_n} = \frac{1}{2}f(\lambda_i)\mu_i^2 + \sigma(\lambda_i),
\quad i = 1,2,\dotsc,n,
\label{eq:47}
\end{equation}
where $\gamma_i \in \mathbb{Z}$ and are such that no two $\gamma_i$ coincide,
and $f,\sigma$ are arbitrary smooth functions. Systems described by separation
relations \eqref{eq:47} are called classical St\"ackel systems.

Consider a St\"ackel system described by a class of irreducible separation
relations given by $n$ copies of the following separation curve (substitution
$\lambda = \lambda_i$, $\mu = \mu_i$ for $i = 1,2,\dotsc,n$ yields $n$
separation relations \eqref{eq:47})
\begin{equation}
H_1 \lambda^{\gamma_1} + H_2 \lambda^{\gamma_2} + \dotsb + H_n =
    \frac{1}{2} f(\lambda) \mu^2 + \sigma(\lambda),
\label{eq:1}
\end{equation}
where $\gamma_1 > \gamma_2 > \dotsb > \gamma_n = 0$, $\gamma_i \in \mathbb{Z}_+$
and $f,\sigma$ are rational functions. Irreducible means, that
the set $\{\gamma_1,\dotsc,\gamma_{n-1}\}$ of integers do not have a common
divisor $\alpha$. Otherwise, separation curve \eqref{eq:1} can be reduced to the
one with $\gamma_i \to \frac{\gamma_i}{\alpha} \in \mathbb{Z}_+$ by a
transformation $\lambda \mapsto \lambda^{\frac{1}{\alpha}}$. The $n$ copies of
\eqref{eq:1} constitute a system of $n$ equations linear in the unknowns
$H_i$ with the solution of the form
\begin{equation}
H_r = \frac{1}{2}(A_r)^{ii} \mu_i^2 + V_r(\lambda)
= \frac{1}{2}(K_r G)^{ii} \mu_i^2 + V_r(\lambda), \quad r = 1,\dotsc,n,
\label{eq:2}
\end{equation}
where $K_r$ are Killing tensors of the metric tensor $G = A_1$ and $K_1 = I$
($K_r$ and $G$ are diagonal in separation coordinates $(\lambda,\mu)$).
Introducing a St\"ackel matrix
\begin{equation}
S_\gamma = \begin{pmatrix}
    \lambda_1^{\gamma_1} &\lambda_1^{\gamma_2}& \cdots & 1 \\
    \vdots & \vdots & & \vdots \\
    \lambda_n^{\gamma_1} & \lambda_n^{\gamma_2}&\cdots & 1
\end{pmatrix}
\label{eq:3}
\end{equation}
separation relations following from \eqref{eq:1} can be written in a
compact form
\begin{equation}
S_\gamma \vec{H} = \vec{U},
\label{eq:48}
\end{equation}
where $\vec{H} = (H_1,\dotsc,H_n)^T$ and $\vec{U} =
(\frac{1}{2}f(\lambda_1) \mu_1^2 + \sigma(\lambda_1),\dotsc,
\frac{1}{2}f(\lambda_n) \mu_n^2 + \sigma(\lambda_n))^T$ is a St\"ackel vector.
It also means that tensor $A_r$ and potential $V_r$ in \eqref{eq:2} can be
expressed as
\begin{equation}
A_r = \diag((S^{-1}_{\gamma})_r^1 f(\lambda_1),\dotsc,
    (S^{-1}_{\gamma})_r^n f(\lambda_n)), \quad
V_r = (S^{-1}_{\gamma})_r^i \sigma(\lambda_i) \quad
r = 1,\dotsc,n,
\label{eq:49}
\end{equation}
and hence
\begin{equation}
H_r = \frac{1}{2} (S^{-1}_{\gamma})_r^i f(\lambda_i) \mu_i^2
    + (S^{-1}_{\gamma})_r^i \sigma(\lambda_i).\label{eq:49a}
\end{equation}
The St\"ackel matrix $S_\gamma$, or equivalently the set $\gamma =
\{\gamma_1,\gamma_2,\dotsc,1\}$, determines a given class of
St\"ackel systems \cite{Blaszak:2009} and we will call it a
\nbr{\gamma}class of classical St\"ackel systems. For a fixed
$S_\gamma$ the metric tensor $G$ is determined by $f(\lambda)$ and
the separable potentials $V_r(\lambda)$ are determined by
$\sigma(\lambda)$. In general metric $G$ is non-flat.

There is one distinguished class of \eqref{eq:1} when $\gamma_k = n - k$, i.e.
\begin{equation}
{H}_1 \lambda^{n-1} + {H}_2 \lambda^{n-2} + \dotsb + {H}_n =
    \frac{1}{2}{f}(\lambda)\mu^2 + {\sigma}(\lambda),
\label{eq:4}
\end{equation}
called Benenti class.

Notice, that all St\"ackel systems \eqref{eq:1} of two degrees of freedom
($n = 2$) are of Benenti type, as the only separation curve \eqref{eq:4} is
irreducible in that case.

For Benenti class, in separation coordinates $(\lambda,\mu)$, the St\"ackel
matrix
\begin{equation}
S = \begin{pmatrix}
    \lambda_1^{n-1} & \cdots & 1 \\
    \vdots & \ddots & \vdots \\
    \lambda_n^{n-1} & \cdots & 1
\end{pmatrix}
\label{eq:51}
\end{equation}
is a Vandermonde matrix and metric tensors are
\begin{equation}
{G}^{ii} = \frac{{f}(\lambda_i)}{\Delta_i}, \quad \Delta_i =
\prod_{k \neq i}(\lambda_i - \lambda_k), \quad i = 1,\dotsc,n.
\label{eq:5}
\end{equation}
All metric tensors \eqref{eq:5} have a common set of Killing tensors
(also diagonal)
\begin{equation}
({K}_r)_i^i = -\frac{\partial \rho_r}{\partial \lambda_i}, \quad r
= 1,\dotsc,n, \label{eq:6}
\end{equation}
where $\rho_r(\lambda)$ are signed symmetric polynomials (Vi\'ete polynomials)
\begin{equation}
\rho_1 = -(\lambda_1 + \dotsb + \lambda_n),\dotsc,
\rho_n = (-1)^n \lambda_1 \lambda_2 \dotsm \lambda_n.
\label{eq:7}
\end{equation}
The matrix
\begin{equation}
F = S^{-1} \Lambda S, \quad \Lambda = \diag(\lambda_1,\dotsc,\lambda_n)
\label{eq:8}
\end{equation}
is a recursion matrix \cite{Blaszak.Marciniak:2012} for basic
potentials ${\sigma}(\lambda) = \lambda^k$
\begin{equation}
{\vec{V}}^{(k)} = F^k {\vec{V}}^{(0)}, \quad k \in \mathbb{Z},
\label{eq:9}
\end{equation}
where ${\vec{V}}^{(k)} = ({V}_1^{(k)},\dotsc,{V}_r^{(k)})^T$,
${V}^{(0)} = (0,\dotsc,0,1)^T$ are separable potentials determined
respectively by ${\sigma}(\lambda) = \lambda^k$ and
${\sigma}(\lambda) = 1$ from separation curve \eqref{eq:4}. In
explicit form
\begin{equation}
F = \begin{pmatrix}
    -\rho_1     & 1      & \cdots & 0 \\
    \vdots      & \vdots & \ddots & \vdots \\
    -\rho_{n-1} & 0      & \cdots & 1 \\
    -\rho_n     & 0      & \cdots & 0
\end{pmatrix}.
\label{eq:10}
\end{equation}
Benenti class of St\"ackel systems contains a sub-class of systems
with flat metrices ${G}$ when
\begin{equation}
{f}(\lambda) = \prod_{k=1}^m (\lambda - \beta_k) =:
{f}_{\text{flat}}(\lambda), \quad m = 0,1,\dotsc,n. \label{eq:11}
\end{equation}

The important fact about St\"ackel systems \eqref{eq:1} is the
existence of a so called St\"ackel transform
\cite{Sergyeyev:2008,Blaszak.Marciniak:2012} relating all of them.
In \cite{Blaszak.Marciniak:2012} it was proved that from a set of
Benenti systems with fixed metric tensor $\bar{G}$ (by fixing
$\bar{f}(\lambda)$), one can construct the rest of St\"ackel
systems \eqref{eq:1}, both from Benenti class as well as from
other classes. The transformation is known as a St\"ackel
transform:
\begin{gather}
\bar{H}_1 \lambda^{n-1} + \bar{H}_2 \lambda^{n-2} + \dotsb +
\bar{H}_n =
    \frac{1}{2}\bar{f}(\lambda)\mu^2 + \bar{\sigma}(\lambda) \nonumber \\
\phantom{\text{ St\"ackel transform}} \bigg\downarrow\text{ St\"ackel transform}
\label{eq:12} \\
H_1 \lambda^{\gamma_1} + H_2 \lambda^{\gamma_2} + \dotsb + H_n =
    \frac{1}{2} f(\lambda) \mu^2 + \sigma(\lambda).
\nonumber
\end{gather}
Explicitly it is given in a matrix form
\begin{equation}
\vec{H} = W_\gamma R(F) \bar{\vec{H}}, \label{eq:13}
\end{equation}
where $\vec{H} = (H_1,\dotsc,H_n)^T$, $\bar{\vec{H}} =
(\bar{H}_1,\dotsc,\bar{H}_n)^T$, $W_\gamma = S_\gamma^{-1} S$,
where $S_\gamma,S$ are respective St\"ackel matrices \eqref{eq:3},
\eqref{eq:51} and $R(F) = f(F) \bar{f}^{-1}(F)$. What is
important, the inverse of the matrix $W_\gamma$ is expressible by
basic potentials $V_i^{(\gamma_j)}$ \eqref{eq:9}
\begin{equation}
(S^{-1} S_\gamma)_{ij} = (W_\gamma^{-1})_{ij} = V_i^{(\gamma_j)}.
\label{eq:14}
\end{equation}

Now, let us choose $\bar{f}(\lambda) =
\bar{f}_{\text{flat}}(\lambda)$ and write $\{\bar{H}_r\}$ in
respective flat coordinates $(x,y)$ (not necessary orthogonal). It
means that all St\"ackel Hamiltonians $\{H_r\}$ of \eqref{eq:1}
can be expressed by a flat coordinates as well, so can be
considered as some quadratic in momenta functions on a phase space
$M = \mathbb{R}^{2n}$.

Consider St\"ackel Hamiltonians \eqref{eq:13} written in a flat
coordinates $(x,y)$ of the metric tensor $\bar{G}$ \eqref{eq:11}
\begin{equation}
H_r = \frac{1}{2} A_r^{ij} y_i y_j + V_r(x), \quad
r = 1,\dotsc,n.
\label{eq:15}
\end{equation}
There are two natural settings for Hamiltonians \eqref{eq:15} as
functions on a phase space $M = T^*\mathcal{Q}$ (a cotangent
bundle to a configuration space $\mathcal{Q}$). We can consider
$\mathcal{Q}$ as two different pseudo-Riemannian spaces. Either
$\mathcal{Q} = (\mathbb{R}^n,\bar{g})$ or $\mathcal{Q} =
(\mathbb{R}^n,g)$, where $\bar{g} = \bar{G}^{-1}$, $g = G^{-1}$,
and $G = A_1$. The second case is natural for classical
separability theory, as then
\begin{equation}
H_r = \frac{1}{2} A_r^{ij} y_i y_j + V_r(x)
= \frac{1}{2}(K_r G)^{ij} y_i y_j + V_r(x),
\label{eq:16}
\end{equation}
$K_1 = I$ and $K_r$ are Killing tensors of the metric $G$,
non-flat in general. Obviously, in the first case, Hamiltonians
\eqref{eq:15} can be written as
\begin{equation}
H_r = \frac{1}{2} A_r^{ij} y_i y_j + V_r(x) = \frac{1}{2}(T_r
\bar{G})^{ij} y_i y_j + V_r(x). \label{eq:17}
\end{equation}
Although tensors $T_r$ are not Killing tensors for the flat metric
$\bar{G}$, but the representation \eqref{eq:17} will be useful for
admissible quantizations of $H_r$.

%%%%%%%%%%%%%%%%%%%%%%%%%%%%%%%%%%%%%%%%%%%%%%%%%%%%%%%%%%%%%%%%%%%%%%%%%%%%%%%%
\section{Minimal quantizations of St\"ackel systems}
\label{sec:3}
Let $(\mathcal{Q},g)$ be a pseudo-Riemannian configuration space and
\begin{equation}
H = \frac{1}{2} A^{ij} p_i p_j + V(q)
\label{eq:18}
\end{equation}
be a function on $T^*\mathcal{Q}$, written in a canonical chart
$(q,p)$ and associated with a symmetric contravariant two-tensor
$A$ on $\mathcal{Q}$. A minimal quantization procedure
\cite{Duval:2005,Blaszak:2013b,Benenti:2002a,Benenti:2002b}
associates with \eqref{eq:18} a self-adjoint linear operator
\begin{equation}
\hat{H} = -\frac{1}{2} \hbar^2 \nabla_i A^{ij} \nabla_j + V(q)
\label{eq:19}
\end{equation}
acting in a Hilbert space $L^2(\mathcal{Q},\omega_g)$ of square integrable
functions defined on the configuration space $\mathcal{Q}$ with respect to the
metric volume form $\omega_g$. By $\nabla_i$ we denote the covariant derivative
with respect to the Levi-Civita connection.

Hence, for St\"ackel Hamiltonians \eqref{eq:15} we can apply either flat or
non-flat minimal quantization related with representations \eqref{eq:16} and
\eqref{eq:17}, respectively. In \cite{Blaszak:2013c} we analyzed the non-flat
case. In the following paper we consider all admissible flat
minimal quantizations and compare them with the non-flat one.

For a non-flat case \eqref{eq:16} the related set of quantum operators is
\begin{equation}
\hat{H}_r = -\frac{1}{2} \hbar^2 \nabla_i A^{ij} \nabla_j + V_r(x), \quad
r = 1,\dotsc,n
\label{eq:20}
\end{equation}
where $\nabla_i$ is the covariant derivative with respect to the connection
generated by metric $g$ and for the flat representation \eqref{eq:17}
respectively
\begin{equation}
\hat{\bar{H}}_r = -\frac{1}{2} \hbar^2 \bar{\nabla}_i A^{ij}
    \bar{\nabla}_j + V_r(x), \quad r = 1,\dotsc,n,
\label{eq:21}
\end{equation}
where $\bar{\nabla}_i$ is the covariant derivative with respect to
the connection generated by a flat metric $\bar{g}$. In order to
investigate a separability of \eqref{eq:20} and \eqref{eq:21}, let
us rewrite the operators in separation coordinates $(\lambda,\mu)$
\cite{Benenti:2002b}
\begin{subequations}
\label{eq:22}
\begin{align}
\hat{H}_r & = -\frac{1}{2}\hbar^2 G^{ii} \left(
    K_r^{(i)} \partial_i^2 + (\partial_i K_r^{(i)})\partial_i
    - K_r^{(i)} \Gamma_i \partial_i\right)
    + V_r(\lambda), \label{eq:22a} \\
\hat{\bar{H}}_r & = -\frac{1}{2}\hbar^2 \bar{G}^{ii} \left(
    T_r^{(i)} \partial_i^2 + (\partial_i T_r^{(i)})\partial_i
    - T_r^{(i)} \bar{\Gamma}_i \partial_i\right)
    + V_r(\lambda), \label{eq:22b}
\end{align}
\end{subequations}
where $\Gamma_i$ ($\bar{\Gamma}_i$) is the contracted Christoffel
symbol defined by $\Gamma_i = g_{il} G^{jk} \Gamma_{jk}^l$ and in
orthogonal coordinates
\begin{equation}
\Gamma_i = \frac{1}{2} \partial_i \ln\abs{G} - \partial_i \ln G^{ii},
\label{eq:23}
\end{equation}
$K_r^{(i)} \equiv (K_r)^i_i$, $T_r^{(i)} \equiv (T_r)^i_i$, and $\partial_i =
\frac{\partial}{\partial \lambda_i}$. As all $K_r$ are Killing tensors for the
metric $G$ so $\partial_i K_r^{(i)} = 0$ \cite{Benenti:2002b}. Thus,
\eqref{eq:22} can be written in the form
\begin{subequations}
\label{eq:24}
\begin{align}
\hat{H}_r & = -\frac{1}{2}\hbar^2 A_r^{ii} \left(
    \partial_i^2 - \Gamma_i \partial_i\right) + V_r(\lambda),
    \label{eq:24a} \\
\hat{\bar{H}}_r & = -\frac{1}{2}\hbar^2 A_r^{ii} \left(
    \partial_i^2 + (\partial_i \ln T_r^{(i)})
    \partial_i - \bar{\Gamma}_i \partial_i\right) + V_r(\lambda).
    \label{eq:24b}
\end{align}
\end{subequations}
A necessary and sufficient condition for separability of operators
\eqref{eq:24a} is a Robertson condition \cite{Benenti:2002a}
\begin{equation*}
\Gamma_i = \Gamma_i(\lambda_i) \Leftrightarrow \partial_j\Gamma_i = 0,
\quad j \neq i,
\end{equation*}
while a necessary and sufficient condition for separability of operators
\eqref{eq:24b} takes the form
\begin{equation*}
\partial_i \ln(T_r^{(i)}) - \bar{\Gamma}_i = \xi_i(\lambda_i) \Leftrightarrow
\partial_j\xi_i = 0, \quad j \neq i.
\end{equation*}
Indeed, if operators \eqref{eq:24} are of the form
\begin{align}
\hat{B}_r & = -\frac{1}{2}\hbar^2 A_r^{ii} \left(
    \partial_i^2 + \gamma_i \partial_i\right) + V_r(\lambda),
    \nonumber \\
 & = -\frac{1}{2}\hbar^2 \left(S^{-1}\right)^i_r f(\lambda_i) \left(
    \partial_i^2 + \gamma_i \partial_i\right) +
    \left(S^{-1}\right)^i_r\sigma(\lambda_i), \quad r = 1,\dotsc,n,
\label{eq:59}
\end{align}
where $\hat{B}_r = \hat{H}_r(\hat{\bar{H}}_r)$ and $\gamma_i =
\gamma_i(\lambda_i)$, then application of St\"ackel matrix $S$ to the system of
eigenvalue problems for \eqref{eq:59}
\begin{equation}
S\begin{pmatrix}
    \hat{B}_1\Psi \\
    \vdots \\
    \hat{B}_n\Psi
\end{pmatrix} = S\begin{pmatrix}
    E_1\Psi \\
    \vdots \\
    E_n\Psi
\end{pmatrix}
\label{eq:60}
\end{equation}
separates \eqref{eq:60} onto $n$ one-dimensional eigenvalue problems
\begin{equation*}
(E_1 \lambda_i^{\gamma_1} + E_2 \lambda_i^{\gamma_2} + \dotsb + E_n)
    \psi_i(\lambda_i) = -\frac{1}{2}\hbar^2
    f(\lambda_i)\left[\frac{\dd^2 \psi_i(\lambda_i)}{\dd \lambda_i^2}
    + \gamma_i(\lambda_i)
    \frac{\dd \psi_i(\lambda_i)}{\dd \lambda_i}\right]
    + \sigma(\lambda_i)\psi_i(\lambda_i),
\end{equation*}
where $\Psi(\lambda_1,\dotsc,\lambda_n) = \prod_{i=1}^n\psi_i(\lambda_i)$. In
the case when $\gamma_i(\lambda_i) = \gamma(\lambda_i)$, $i = 1,\dotsc,n$, we
have $n$ copies of one-dimensional eigenvalue problem
\begin{equation*}
(E_1 \lambda^{\gamma_1} + E_2 \lambda^{\gamma_2} + \dotsb + E_n)
    \psi(\lambda) = -\frac{1}{2}\hbar^2
    f(\lambda)\left[\frac{\dd^2 \psi(\lambda)}{\dd \lambda^2}
    + \gamma(\lambda)
    \frac{\dd \psi(\lambda)}{\dd \lambda}\right]
    + \sigma(\lambda)\psi(\lambda),
\end{equation*}
where $\Psi(\lambda_1,\dotsc,\lambda_n) = \prod_{i=1}^n\psi(\lambda_i)$.

%%%%%%%%%%%%%%%%%%%%%%%%%%%%%%%%%%%%%%%%%%%%%%%%%%%%%%%%%%%%%%%%%%%%%%%%%%%%%%%%
\section{Minimal flat quantization of Benenti class}
\label{sec:4}
First, let us analyze the case of two quantizations inside the Benenti class,
where in \eqref{eq:13} $W_\gamma = I$. Assume that $\{\bar{H}_r\}$ is a Benenti
system with a flat metric generated by $\bar{f}_{\text{flat}}(\lambda)$. Then,
any other Benenti system $\{H_r\}$ is given by
\begin{equation}
\vec{H} = R(F)\bar{\vec{H}}, \quad
R(F) = f(F)\bar{f}_{\text{flat}}^{-1}(F)
\label{eq:25}
\end{equation}
and separation curves for $\{\bar{H}_r\}$ and $\{H_r\}$ are
\begin{gather}
\bar{H}_1 \lambda^{n-1} + \bar{H}_2 \lambda^{n-2} + \dotsb + \bar{H}_n =
    \frac{1}{2}\bar{f}_{\text{flat}}(\lambda)\mu^2 + \bar{\sigma}(\lambda)
    \nonumber \\
\phantom{R(F)\ } \bigg\downarrow \ R(F) \nonumber \\
H_1 \lambda^{n-1} + H_2 \lambda^{n-2} + \dotsb + H_n =
    \frac{1}{2}f(\lambda)\mu^2 + \sigma(\lambda),
\label{eq:26}
\end{gather}
$\sigma(\lambda) = R(\lambda)\bar{\sigma}(\lambda) =
f(\lambda)\bar{\sigma}(\lambda)/\bar{f}_{\text{flat}}(\lambda)$.
The relation \eqref{eq:26} follows from the following relations which hold in
separation coordinates:
\begin{equation}
A_r = \sum_k R(F)_{rk} \bar{A}_k = R(\Lambda)\bar{A}_r,
\label{eq:61}
\end{equation}
\begin{equation}
\vec{V} = R(F)\bar{\vec{V}} = S^{-1}R(\Lambda)\bar{\vec{\sigma}}(\lambda), \quad
\bar{\vec{\sigma}}(\lambda) =
    (\bar{\sigma}(\lambda_1),\dotsc,\bar{\sigma}(\lambda_n))^T.
\label{eq:62}
\end{equation}
Indeed, for rational $f(\lambda)$ \eqref{eq:61} follows from the fact that it
is fulfilled for $R(F) = F - \beta I$ and $R(F) = (F - \beta I)^{-1}$. To prove
\eqref{eq:62} observe that $\bar{\vec{V}} = S^{-1}\bar{\vec{\sigma}}(\lambda)$
and $R(F) = S^{-1}R(\Lambda)S$. Hence $\vec{V} = R(F)S^{-1}
\bar{\vec{\sigma}}(\lambda) = S^{-1}R(\Lambda)\bar{\vec{\sigma}}(\lambda)$.

Now, let us go back to operators \eqref{eq:24}. As for metric \eqref{eq:5} from
Benenti class $\Gamma_i = -\frac{1}{2}
\frac{\partial_i f(\lambda_i)}{f(\lambda_i)}$ and as follows from \eqref{eq:61}
$T_r^{(i)} = R(\lambda_i) \bar{K}_r^{(i)}$ then, using the relation
\eqref{eq:49}, we have
\begin{subequations}
\label{eq:28}
\begin{align}
\hat{H}_r & = -\frac{1}{2}\hbar^2 (S^{-1})_r^i
    \left(f(\lambda_i)\partial_i^2
    + \frac{1}{2}\frac{\dd f(\lambda_i)}{\dd\lambda_i}\partial_i\right)
    + (S^{-1})_r^i \sigma(\lambda_i), \label{eq:28a} \\
\hat{\bar{H}}_r & = -\frac{1}{2}\hbar^2 (S^{-1})_r^i
    \left[f(\lambda_i)\partial_i^2
    + \left(\frac{\dd f(\lambda_i)}{\dd\lambda_i}
    - \frac{1}{2} \frac{f(\lambda_i)}{\bar{f}_{\text{flat}}(\lambda_i)}
    \frac{\dd\bar{f}_{\text{flat}}(\lambda_i)}{\dd\lambda_i}\right)\partial_i
    \right]
    + (S^{-1})_r^i \sigma(\lambda_i), \label{eq:28b}
\end{align}
\end{subequations}
so equations \eqref{eq:28} take the form \eqref{eq:59} with
$\gamma(\lambda_i) = \frac{1}{2}\frac{\dd f(\lambda_i)}{\dd\lambda_i}$ in the
case of Eq.~\eqref{eq:28a} and $\gamma(\lambda_i) =
\frac{\dd f(\lambda_i)}{\dd\lambda_i} -
\frac{1}{2} \frac{f(\lambda_i)}{\bar{f}_{\text{flat}}(\lambda_i)}
\frac{\dd\bar{f}_{\text{flat}}(\lambda_i)}{\dd\lambda_i}$ in the
case of Eq.~\eqref{eq:28b}. As a consequence all operators
$\{\hat{H}_r\}$ as well as $\{\hat{\bar{H}}_r\}$ have common
eigenfunctions:
\begin{equation}
\hat{H}_r \Psi = E_r \Psi, \quad
\hat{\bar{H}}_r \bar{\Psi} = \bar{E}_r \bar{\Psi}, \quad r = 1,\dotsc,n,
\label{eq:29}
\end{equation}
where $\Psi(\lambda_1,\dotsc,\lambda_n) = \prod_{k=1}^n
\psi(\lambda_k)$, $\bar{\Psi}(\lambda_1,\dotsc,\lambda_n) =
\prod_{k=1}^n \bar{\psi}(\lambda_k)$, and $\psi(\lambda_k)$ and
$\bar{\psi}(\lambda_k)$ are $n$ copies of one-dimensional
eigenvalue problems
\begin{subequations}
\label{eq:30}
\begin{align}
(E_1 \lambda^{n-1} + E_2 \lambda^{n-2} + \dotsb + E_n)
    \psi(\lambda) & = -\frac{1}{2}\hbar^2
    \left(f(\lambda)\frac{\dd^2 \psi(\lambda)}{\dd \lambda^2}
    + \frac{1}{2}\frac{\dd f(\lambda)}{\dd\lambda}
    \frac{\dd \psi(\lambda)}{\dd \lambda}\right)
    + \sigma(\lambda)\psi(\lambda), \label{eq:30a} \\
(\bar{E}_1 \lambda^{n-1} + \bar{E}_2 \lambda^{n-2} + \dotsb + \bar{E}_n)
    \bar{\psi}(\lambda) & = -\frac{1}{2}\hbar^2
    \left[f(\lambda)\frac{\dd^2 \bar{\psi}(\lambda)}{\dd \lambda^2}
    + \left(\frac{\dd f(\lambda)}{\dd\lambda}
    - \frac{1}{2} \frac{f(\lambda)}{\bar{f}_{\text{flat}}(\lambda)}
    \frac{\dd \bar{f}_{\text{flat}}(\lambda)}{\dd \lambda}\right)
    \frac{\dd \bar{\psi}(\lambda)}{\dd \lambda}\right]
    + \sigma(\lambda)\bar{\psi}(\lambda).
\label{eq:30b}
\end{align}
\end{subequations}
Equations \eqref{eq:30a} and \eqref{eq:30b} represent the non-flat and flat
minimal quantizations of separation curve \eqref{eq:26}. Moreover,
\begin{equation}
[\hat{H}_r,\hat{H}_s] = 0, \quad [\hat{\bar{H}}_r,\hat{\bar{H}}_s] = 0.
\label{eq:31}
\end{equation}
The first set of commutation relations was proved in \cite{Blaszak:2013c} and
follows from the fulfillment of the pre-Robertson condition \cite{Benenti:2002b}
\begin{equation*}
\partial_i^2\Gamma_i - \Gamma_j\partial_i\Gamma_j = 0, \quad i \neq j
\end{equation*}
for $\Gamma_i = -\frac{1}{2}\partial_i\ln f(\lambda_i)$. The second set of
commutation relations follows from the analog of the pre-Robertson condition
\begin{equation}
\partial_i^2\gamma_i - \gamma_j\partial_i\gamma_j = 0, \quad i \neq j,
\label{eq:63}
\end{equation}
where
\begin{equation*}
\gamma_i = \bar{\Gamma}_i - \partial_i\ln R(\lambda_i)
= -\frac{1}{2}\partial_i\ln f_{\text{flat}}(\lambda_i)
    - \partial_i\ln R(\lambda_i).
\end{equation*}
The condition \eqref{eq:63} can be obtain repeating the procedure from
\cite{Benenti:2002b} (Section V) under substitution $K_r^{(i)} \rightarrow
R(\Lambda)K_r^{(i)}$.

Summarizing that part, we proved that for any classical Benenti system, there
exists an \nbr{n}parameter family \eqref{eq:11} of minimal flat quantizations,
which preserves quantum separability.

%%%%%%%%%%%%%%%%%%%%%%%%%%%%%%%%%%%%%%%%%%%%%%%%%%%%%%%%%%%%%%%%%%%%%%%%%%%%%%%%
\section{Minimal flat quantization for arbitrary $\gamma$-class}
\label{sec:5}
Let us consider the case $R = 1$. Then,
\begin{equation}
\vec{H} = W_\gamma \bar{\vec{H}}, \label{eq:32}
\end{equation}
where separation curves for ${\bar{H}_r}$ and ${H_r}$ are
\begin{gather}
\bar{H}_1 \lambda^{n-1} + \bar{H}_2 \lambda^{n-2} + \dotsb +
\bar{H}_n =
    \frac{1}{2}\bar{f}_{\text{flat}}(\lambda)\mu^2 + \bar{\sigma}(\lambda)
    \nonumber \\
\phantom{W_\gamma\ } \bigg\downarrow \ W_\gamma \label{eq:33} \\
H_1 \lambda^{\gamma_1} + H_2 \lambda^{\gamma_2} + \dotsb + H_n =
    \frac{1}{2}\bar{f}_{\text{flat}}(\lambda)\mu^2 + \bar{\sigma}(\lambda),
    \nonumber
\end{gather}
and Hamiltonian operators for non-flat and flat minimal quantizations of ${H_i}$
are of the form \eqref{eq:24}. $\Gamma_i$ in \eqref{eq:24a} is a reduced
Christoffel symbol for a metric tensor $G$, and it was proved in
\cite{Blaszak:2013c} that for arbitrary \nbr{\gamma}class
$\partial_j \Gamma_i \neq 0$, $j \neq i$ and we loose a separability. In
operator $\hat{\bar{H}}_r$ from \eqref{eq:24b} $\bar{\Gamma}_i = -\frac{1}{2}
\frac{\partial_i\bar{f}_{\text{flat}}(\lambda_i)}{
\bar{f}_{\text{flat}}(\lambda_i)}$, hence does not depend on
$\lambda_j \neq \lambda_i$, so we have to analyze only the term
$\partial_i T_r^{(i)}/T_r^{(i)}$. A very useful form of $T_r^{(i)}$ was derived
in \cite{Blaszak:2005}. Consider polynomial
$P = \sum_{r=1}^n H_r \lambda^{\gamma_r}$ from separation curve \eqref{eq:33}.
Its order is $\gamma_1$ which we denote by $\gamma_1 = n + k - 1$. Notice that
for $k = 0$ we are in the Benenti class. There is $k$ missing monomials
$\lambda^{n + k - n_i}$ in polynomial $P$, enumerated by $(n_1,\dotsc,n_k)$. For
example, if $P = H_1 \lambda^4 + H_2 \lambda + H_3$, then $n = 3$, $k = 2$,
$n_1 = 2$, $n_2 = 3$. In \cite{Blaszak:2005} was proved that
\begin{equation}
T_r^{(i)} = \frac{1}{\varphi} \chi_r^{(i)},
\label{eq:34}
\end{equation}
where $\chi_r^{(i)}$ is \nbr{\lambda_i}independent and
\begin{equation}
\varphi = \det\begin{pmatrix}
    \rho_{n_1 - 1} & \cdots & \rho_{n_1 - k} \\
    \vdots         & \ddots & \vdots \\
    \rho_{n_k - 1} & \cdots & \rho_{n_k - k}
\end{pmatrix}
\label{eq:35}
\end{equation}
where $\rho_0 = 1$, $\rho_m = 0$ for $m > n$ and $m < 0$, and remaining $\rho_m$
are given by \eqref{eq:7}. Hence, \eqref{eq:24b} takes the form
\begin{equation}
\hat{\bar{H}}_r = -\frac{1}{2}\hbar^2 A_r^{ii}
    \left(\partial_i^2 - \left(\frac{\partial_i \varphi}{\varphi} - \frac{1}{2}
    \frac{\partial_i \bar{f}_{\text{flat}}}{\bar{f}_{\text{flat}}}\right)
    \partial_i\right) + V_r(\lambda).
\label{eq:36}
\end{equation}
It can be proved that for any $\varphi$,
$\partial_j\left(\frac{\partial_i \varphi}{\varphi}\right) \neq 0$ for
$j \neq i$. As a result, all admissible flat minimal quantizations of a
non-Benenti \nbr{\gamma}class destroy a quantum separability.

%%%%%%%%%%%%%%%%%%%%%%%%%%%%%%%%%%%%%%%%%%%%%%%%%%%%%%%%%%%%%%%%%%%%%%%%%%%%%%%%
\section{Separable deformations of St\"ackel Hamiltonians}
\label{sec:6}
In order to make all Hamiltonians $\hat{\bar{H}}_r$ separable, we need to get
rid of the terms
\begin{equation}
\frac{1}{2}\hbar^2\left(A_r^{ii}\frac{\partial_i\varphi}{\varphi}\right)
    \partial_i
\label{eq:37}
\end{equation}
from \eqref{eq:36}. Terms \eqref{eq:37} are generated by appropriate linear in
momenta terms in Hamiltonians $H_r$. Define a vector field $u_r$ with components
\begin{equation}
u_r^i = A^{ii}_r \frac{\partial_i \varphi}{\varphi}
\label{eq:38}
\end{equation}
in separation coordinates. Then, consider a deformed Hamiltonians in flat
coordinates
\begin{equation}
H_r(\hbar) = \frac{1}{2}A_r^{ij} y_i y_j - \frac{1}{2}i\hbar u_r^i(x) y_i
    + V_r(x) + \frac{1}{4}\hbar^2 w_r(x),
\label{eq:39}
\end{equation}
where $w_r = \sum_i \frac{\partial u_r^i}{\partial x_i}$. Appropriate quantum
operator in flat minimal quantization takes a form
\begin{equation}
\hat{\bar{H}}_r = -\frac{1}{2}\hbar^2 \bar{\nabla}_i A_r^{ij}
\bar{\nabla}_j
    - \frac{1}{4}\hbar^2(\bar{\nabla}_i u_r^i + u_r^i \bar{\nabla}_i)
    + \frac{1}{4}\hbar^2 w_r(x) + V_r(x)
\label{eq:40}
\end{equation}
and in separation coordinates
\begin{equation}
\hat{\bar{H}}_r = -\frac{1}{2}\hbar^2 A^{ii} \left[\partial_i^2
    + \frac{1}{2}(\partial_i \ln{\bar{f}_{\text{flat}}(\lambda_i)})
    \partial_i\right] + V_r(\lambda).
\label{eq:41}
\end{equation}
Hence all $\hat{\bar{H}}_r$ separate to a single one-dimensional
eigenvalue problem:
\begin{equation}
(E_1 \lambda^{\gamma_1} + E_2 \lambda^{\gamma_2} + \dotsb + E_n)
    \bar{\psi}(\lambda) = -\frac{1}{2}\hbar^2
    \left(\bar{f}_{\text{flat}}(\lambda)
    \frac{\dd^2\bar{\psi}(\lambda)}{\dd\lambda^2}
    + \frac{1}{2}\frac{\dd \bar{f}_{\text{flat}}(\lambda)}{\dd \lambda}
    \frac{\dd \bar{\psi}(\lambda)}{\dd \lambda}\right)
    + \bar{\sigma}(\lambda)\bar{\psi}(\lambda).
\label{eq:42}
\end{equation}
Nevertheless $\hat{\bar{H}}_r$ are not Hermitian anymore, since the extra terms
$-\frac{1}{4}\hbar^2(\bar{\nabla}_i u_r^i + u_r^i \bar{\nabla}_i)$ are
anti-Hermitian operators in a Hilbert space $L^2(\mathcal{Q},\omega_g)$.

%%%%%%%%%%%%%%%%%%%%%%%%%%%%%%%%%%%%%%%%%%%%%%%%%%%%%%%%%%%%%%%%%%%%%%%%%%%%%%%%
\section{Examples}
\label{sec:7}
As first example let us consider a pseudo-Euclidean space $E^3$ with signature
$(++-)$ and flat non-orthogonal coordinates $(x_1,x_1,x_3)$ such that
\begin{equation}
\bar{g} = \begin{pmatrix}
    0 & 0 & 1 \\
    0 & 1 & 0 \\
    1 & 0 & 0
\end{pmatrix}.
\label{eq:50}
\end{equation}
Then, consider the following St\"ackel geodesic system on $T^*E^3$
\begin{align*}
\bar{h}_1 & = \bar{G}^{ij}y_i y_j = y_1 y_3 + \frac{1}{2}y_2^2, \nonumber \\
\bar{h}_2 & = (\bar{K}_2\bar{G})^{ij}y_i y_j = \frac{1}{8}x_1^2
y_1^2 - \frac{1}{4}x_1 x_3 y_2^2 + \frac{1}{8}x_3^2 y_3^2
    + \left(\frac{1}{4}x_1 x_2 + 1\right)y_1 y_2
    - \frac{1}{4}\left(x_1 x_3 + x_2^2\right)y_1 y_3
    - \frac{1}{4}x_2 x_3 y_2 y_3, \nonumber \\
\bar{h}_3 & = (\bar{K}_3\bar{G})^{ij}y_i y_j =
\left(\frac{1}{4}x_1x_2 + \frac{1}{2}\right)y_1^2 - \frac{1}{4}x_1
x_3 y_1 y_2
    - \frac{1}{4}x_2 x_3 y_1 y_3 + \frac{1}{4}x_3^2 y_2 y_3.
\end{align*}
One can check that $\{\bar{h}_i,\bar{h}_j\} = 0.$ The
transformation to separation coordinates $(\lambda,\mu)$ is
generated by \cite{Blaszak:2007}
\begin{gather}
\lambda_1 + \lambda_2 + \lambda_3 = \frac{1}{2}x_1 x_3 + \frac{1}{4}x_2^2,
\nonumber \\
\lambda_1\lambda_2 + \lambda_1\lambda_3 + \lambda_2\lambda_3 =
    -\frac{1}{2}x_2 x_3,
\label{eq:52} \\
\lambda_1\lambda_2\lambda_3 = \frac{1}{4}x_3^2 \nonumber
\end{gather}
and the related separation curve is
\begin{equation*}
\bar{h}_1\lambda^2 + \bar{h}_2\lambda + \bar{h}_3 =
    \frac{1}{2}\lambda^3\mu^2.
\end{equation*}
operator $F$ \eqref{eq:10} in \nbr{x}coordinates is
\begin{equation*}
F = \begin{pmatrix}
    \frac{1}{2}x_1 x_3 + \frac{1}{4}x_2^2 & 1 & 0 \\
    \frac{1}{2}x_2 x_3                    & 0 & 1 \\
    \frac{1}{4}x_3^2                      & 0 & 0
\end{pmatrix},
\end{equation*}
so separable potentials $\bar{V}^{(k)}_r$ are given by \eqref{eq:9}. For
example, the first nontrivial potential is
\begin{equation*}
\bar{\vec{V}}^{(3)} = F^3\bar{\vec{V}}^{(0)} = \begin{pmatrix}
    \frac{1}{2}x_1 x_3 + \frac{1}{4}x_2^2 \\
    \frac{1}{2}x_2 x_3 \\
    \frac{1}{4}x_3^2
\end{pmatrix}
\end{equation*}
and separation curve for Hamiltonians $\bar{H}_i = \bar{h}_i + \bar{V}^{(k)}_i$,
$i = 1,2,3$, takes the form
\begin{equation*}
\bar{H}_1\lambda^2 + \bar{H}_2\lambda + \bar{H}_3 =
    \frac{1}{2}\lambda^3\mu^2 + \lambda^k.
\end{equation*}
Now, let us consider the following St\"ackel transform
\begin{gather}
\bar{H}_1 \lambda^{2} + \bar{H}_2 \lambda + \bar{H}_3 =
    \frac{1}{2}\lambda^3\mu^2 + \lambda^{r-s+3} \nonumber \\
\phantom{R(F)=F^{s-3}\ } \bigg\downarrow \ R(F)=F^{s-3} \nonumber \\
H_1 \lambda^{2} + H_2 \lambda + H_3 =
    \frac{1}{2}\lambda^s\mu^2 + \lambda^r \label{eq:53}
\end{gather}
so, $\vec{H} = F^{s-3}\bar{\vec{H}}$ and in particular, for $s = 4$ and $r = 4$,
we have for $H_i = h_i + V_i^{(4)}$
\begin{align*}
h_1 & = \frac{1}{8}x_1^2 y_1^2 + \frac{1}{8}x_2^2 y_2^2 + \frac{1}{8}x_3^2 y_3^2
    + \left(\frac{1}{4}x_1 x_2 + 1\right)y_1 y_2 + \frac{1}{4}x_1 x_3 y_1 y_3
    + \frac{1}{4}x_2 x_3 y_2 y_3, \nonumber \\
h_2 & = \left(\frac{1}{4}x_1 x_2 + \frac{1}{2}\right)y_1^2
    + \frac{1}{4}x_2 x_3 y_2^2 - \frac{1}{4}x_1 x_3 y_1 y_2
    + \frac{1}{4}x_2 x_3 y_1 y_3 + \frac{1}{4}x_3^2 y_2 y_3, \nonumber \\
h_3 & = \frac{1}{4}x_3^2 y_1 y_3 + \frac{1}{8}x_3^2 y_2^2
\end{align*}
and
\begin{align}
V_1^{(4)} & = \frac{1}{4}x_1^2 x_3^2 \frac{1}{4}x_1 x_2^2 x_3
    + \frac{1}{16}x_2^4 + \frac{1}{2}x_2 x_3, \nonumber \\
V_2^{(4)} & = \frac{1}{4}x_1 x_2 x_3^2 + \frac{1}{8}x_2^3 x_3
    + \frac{1}{4}x_2 x_3,\nonumber \\
V_3^{(4)} & = \frac{1}{16}x_3^2\left(2x_1 x_3 + x_2^2\right).
\label{eq:54}
\end{align}
Of course, again canonical transformation generated by \eqref{eq:52} is a
transformation to separation coordinates, with separation curve \eqref{eq:53}
and $s = r = 4$.

As was considered in previous sections, we have two natural minimal
quantizations. One, the flat minimal quantization expressed by Levi-Civita
connection of metric $\bar{g}$ \eqref{eq:50} and second, expressed by
Levi-Civita connection of metric tensor $g = G^{-1}$, where
\begin{equation*}
G = \begin{pmatrix}
    \frac{1}{4}x_1^2       & \frac{1}{4}x_1 x_2 + 1 & \frac{1}{4}x_1 x_3 \\
    \frac{1}{4}x_1 x_2 + 1 & \frac{1}{4}x_2^2       & \frac{1}{4}x_2 x_3 \\
    \frac{1}{4}x_1 x_3     & \frac{1}{4}x_1 x_3     & \frac{1}{4}x_3^2
\end{pmatrix}
\end{equation*}
is generated by $h_1 = \frac{1}{2}G^{ij}y_i y_j$. As $g$ has constant Ricci
scalar $R_S = \frac{3}{2}$, the second admissible minimal quantization is
non-flat.

In flat quantization, related to metric tensor $\bar{g}$ Christoffel symbols
vanish and quantum operators $\hat{\bar{H}}_r$ related to classical Hamiltonian
functions $H_r$ are
\begin{align}
\hat{\bar{H}}_1 & = -\hbar^2\left[\frac{1}{8}\left(x_1^2\partial_1^2
    + x_2^2\partial_2^2 + x_3^2\partial_3^2\right)
    + \left(\frac{1}{4}x_1 x_2 + 1\right)\partial_1\partial_2
    + \frac{1}{4}x_1 x_3\partial_1\partial_3
    + \frac{1}{4}x_2 x_3\partial_2\partial_3
    + \frac{1}{2}\left(x_1\partial_1 + x_2\partial_2 + x_3\partial_3\right)
    \right] \nonumber \\
& \quad {} + V_1^{(r)}, \nonumber \\
\hat{\bar{H}}_2 & = -\hbar^2\left[\left(\frac{1}{4}x_1 x_2
    + \frac{1}{2}\right)\partial_1^2 + \frac{1}{4}x_2 x_3\partial_2^2
    - \frac{1}{4}x_1 x_3\partial_1\partial_2
    + \frac{1}{4}x_2 x_3\partial_1\partial_3
    + \frac{1}{4}x_3^2\partial_2\partial_3 + \frac{3}{8}x_2\partial_1
    + \frac{3}{8}x_3\partial_2\right]
    + V_2^{(r)}, \nonumber \\
\hat{\bar{H}}_3 & = -\hbar^2\left[\frac{1}{4}x_3^2\partial_1\partial_3
    + \frac{1}{8}x_3^2\partial_2^2 + \frac{1}{4}x_3\partial_1\right]
    + V_3^{(r)}.
\label{eq:55}
\end{align}
Obviously these operators are Hermitian in $L^2(\mathcal{Q},\omega_g)$.
Substituting $r = 4$ \eqref{eq:54} one can check directly the commutativity of
operators $\hat{\bar{H}}_r$ \eqref{eq:55}.

In $(\lambda,\mu)$ coordinates eigenvalue problems \eqref{eq:29} reduce to three
copies of one-dimensional eigenvalue problem
\begin{equation*}
(\bar{E}_1 \lambda^2 + \bar{E}_2 \lambda + \bar{E}_3)\bar{\psi}(\lambda)
    = -\frac{1}{2}\hbar^2\left[\lambda^s\frac{\dd^2 \bar{\psi}}{\dd \lambda^2}
    + \left(s - \frac{3}{2}\right)\lambda^{s-1}
    \frac{\dd \bar{\psi}}{\dd \lambda}\right]
    + \lambda^r\bar{\psi}
\end{equation*}
for operators $\hat{\bar{H}}_r$ of minimal flat quantization and
\begin{equation*}
(E_1 \lambda^2 + E_2 \lambda + E_3) \psi(\lambda) = -\frac{1}{2}\hbar^2
    \left[\lambda^s\frac{\dd^2 \psi}{\dd \lambda^2}
    + \frac{1}{2}\lambda^{s-1}\frac{\dd \psi}{\dd \lambda}\right]
    + \lambda^r\psi,
\end{equation*}
for operators $\hat{H}_r$ of minimal non-flat quantization with
$f(\lambda) = \lambda^s$.

As our second example let us consider again a pseudo-Euclidean space $E^3$
with signature $(++-)$ and flat, non-orthogonal coordinates $(x_1,x_1,x_3)$ such
that
\begin{equation}
\bar{g} = \begin{pmatrix}
    0 & 1 & 0 \\
    1 & 0 & 0 \\
    0 & 0 & 1
\end{pmatrix}.
\label{eq:56}
\end{equation}
Then, consider the following St\"ackel geodesic system on $T^*E^3$
\begin{align*}
\bar{h}_1 & = \bar{G}^{ij}y_i y_j = y_1 y_2 + \frac{1}{2}y_3^2, \nonumber \\
\bar{h}_2 & = (\bar{K}_2\bar{G})^{ij}y_i y_j = \frac{1}{2}y_1^2 -
\frac{1}{2}x_2 y_2^2 + \frac{1}{2}x_1 y_3^2
    + \frac{1}{2}x_1 y_1 y_2 - \frac{1}{2}x_3 y_2 y_3, \nonumber \\
\bar{h}_3 & = (\bar{K}_3\bar{G})^{ij}y_i y_j = \frac{1}{8}x_3^2
y_2^2 + \left(\frac{1}{8}x_1^2 + \frac{1}{2}x_2\right)y_3^2
    - \frac{1}{2}x_3 y_1 y_3 - \frac{1}{4}x_1 x_3 y_2 y_3.
\end{align*}
One can check that $\{\bar{h}_i,\bar{h}_j\}=0.$ The transformation
to separation coordinates $(\lambda,\mu)$ is generated by
\cite{Blaszak:2007}
\begin{gather}
\lambda_1 + \lambda_2 + \lambda_3 = -x_1, \nonumber \\
\lambda_1\lambda_2 + \lambda_1\lambda_3 + \lambda_2\lambda_3 =
    x_2 + \frac{1}{4}x_1^2,
\label{eq:57} \\
\lambda_1\lambda_2\lambda_3 = \frac{1}{4}x_3^2. \nonumber
\end{gather}
The related separation curve is
\begin{equation*}
\bar{h}_1\lambda^2 + \bar{h}_2\lambda + \bar{h}_3 =
    \frac{1}{2}\lambda\mu^2,
\end{equation*}
operator $F$ \eqref{eq:10} in \nbr{x}coordinates takes the form
\begin{equation*}
F = \begin{pmatrix}
    -x_1                    & 1 & 0 \\
    -x_2 - \frac{1}{4}x_1^2 & 0 & 1 \\
    \frac{1}{4}x_3^2        & 0 & 0
\end{pmatrix}
\end{equation*}
so, separable potentials $\bar{V}^{(k)}_r$ are given by \eqref{eq:9}. For
example, the $\bar{V}^{(4)}$ potential and separation curve for Hamiltonians
$\bar{H}_i = \bar{h}_i + \bar{V}^{(4)}_i$ are
\begin{gather*}
\bar{\vec{V}}^{(4)} = F^4\bar{\vec{V}}^{(0)} = \begin{pmatrix}
    \frac{3}{4}x_1^2 - x_2 \\
    \frac{1}{4}x_1^3 + x_1 x_2 + \frac{1}{4}x_3^2 \\
    -\frac{1}{4}x_1 x_3^2
\end{pmatrix} \\
\bar{H}_1\lambda^2 + \bar{H}_2\lambda + \bar{H}_3 =
    \frac{1}{2}\lambda\mu^2 + \lambda^4.
\end{gather*}
First, let us consider the following St\"ackel transform
\begin{gather}
\bar{H}_1 \lambda^2 + \bar{H}_2 \lambda + \bar{H}_3 =
    \frac{1}{2}\lambda\mu^2 + \lambda^4 \nonumber \\
\phantom{W_{\gamma}\ } \bigg\downarrow \ W_{\gamma} \nonumber \\
H_1 \lambda^3 + H_2 \lambda + H_3 =
    \frac{1}{2}\lambda\mu^2 + \lambda^4, \label{eq:58}
\end{gather}
where $\gamma = (3,1,0)$ and from \eqref{eq:14}
\begin{equation*}
W_{\gamma} = \begin{pmatrix}
    -\frac{1}{x_1}                       & 0 & 0 \\
    -\frac{1}{4}\frac{x_1^2 + 4x_2}{x_1} & 1 & 0 \\
    \frac{1}{4}\frac{x_3^2}{x_1}         & 0 & 1
\end{pmatrix}.
\end{equation*}
Then, according to \eqref{eq:32}
\begin{align*}
H_1 & = -\frac{1}{x_1}y_1 y_2 - \frac{1}{2}\frac{1}{x_1}y_3^2 - \frac{3}{4}x_1
    + \frac{x_2}{x_1}, \nonumber \\
H_2 & = \frac{1}{2}y_1^2 - \frac{1}{2}x_2 y_2^2
    + \frac{1}{8}\left(3x_1 - 4\frac{x_2}{x_1}\right)y_3^2
    + \frac{1}{4}\left(x_1 - 4\frac{x_2}{x_1}\right)y_1 y_2
    - \frac{1}{2}x_3 y_2 y_3 + \frac{1}{16}x_1^3 + \frac{1}{2}x_1 x_2
    + \frac{1}{4}x_3^2 + \frac{x_2^2}{x_1}, \nonumber \\
H_3 & = \frac{1}{8}x_3^2 y_2^2 + \frac{1}{8}\left(x_1^2 + 4x_2
    + \frac{x_3^2}{x_1}\right)y_3^2 + \frac{1}{4}\frac{x_3^2}{x_1}y_1 y_2
    - \frac{1}{2}x_3 y_1 y_3 - \frac{1}{4}x_1 x_3 y_2 y_3
    - \frac{1}{16}x_1 x_3^2 - \frac{1}{4}\frac{x_2 x_3^2}{x_1},
\end{align*}
where
\begin{gather*}
A_1 = \begin{pmatrix}
    0 & -\frac{1}{x_1} & 0 \\
    -\frac{1}{x_1} & 0 & 0 \\
    0 & 0 & -\frac{1}{x_1}
\end{pmatrix}, \quad
A_2 = \begin{pmatrix}
    1 & \frac{1}{4}x_1 - \frac{x_2}{x_1} & 0 \\
    \frac{1}{4}x_1 - \frac{x_2}{x_1} & -x_2 & -\frac{1}{2}x_3 \\
    0 & -\frac{1}{2}x_3 & \frac{3}{4}x_1 - \frac{x_2}{x_1}
\end{pmatrix}, \\
A_3 = \begin{pmatrix}
    0 & \frac{1}{4}\frac{x_3^2}{x_1} & -\frac{1}{2}x_3 \\
    \frac{1}{4}\frac{x_3^2}{x_1} & \frac{1}{4}x_3^2 & -\frac{1}{4}x_1 x_3 \\
   -\frac{1}{2}x_3 & -\frac{1}{4}x_1 x_3
   & \frac{1}{4}x_1^2 + x_2 + \frac{1}{4}\frac{x_3^2}{x_1}
\end{pmatrix}.
\end{gather*}
Of course, again canonical transformation generated by \eqref{eq:57} is a
transformation to separation coordinates, with separation curve \eqref{eq:58}.

We have two natural minimal quantizations. One, the flat minimal
quantization expressed by Levi-Civita connection of metric
$\bar{g}$ \eqref{eq:56} and second, expressed by Levi-Civita
connection of metric tensor $g = G^{-1}$, where $G = A_1$.

In $(\lambda,\mu)$ coordinates Hamiltonian operators for non-flat and flat
minimal quantizations are given respective by \eqref{eq:24a} and \eqref{eq:36}.
As
\begin{equation*}
\Gamma_i = -\frac{1}{2}\left(\frac{1}{\lambda_i}
    + \frac{1}{\lambda_1 + \lambda_2 + \lambda_3}\right), \quad
\frac{\partial_i\varphi}{\varphi} = \frac{1}{\lambda_1 + \lambda_2 + \lambda_3},
\end{equation*}
hence both quantizations are non-separable.

The deformation \eqref{eq:39} of classical Hamiltonians, with respective vector
fields
\begin{equation*}
u_1 = \left(0,-\frac{1}{x_1^2},0\right), \quad
u_2 = \left(\frac{1}{x_1},\frac{1}{4} - \frac{x_2}{x_1^2},0\right), \quad
u_3 = \left(0,\frac{1}{4}\frac{x_3^2}{x_1^2},-\frac{x_3}{x_1}\right),
\end{equation*}
leads to commuting (non-Hermitian) operators \eqref{eq:40} and the following
one-dimensional eigenvalue problem
\begin{equation*}
(E_1 \lambda^3 + E_2 \lambda + E_3)\bar{\psi}(\lambda) =
-\frac{1}{2}\hbar^2
    \left(\lambda\frac{\dd^2 \bar{\psi}}{\dd \lambda^2}
    + \frac{1}{2} \frac{\dd \bar{\psi}}{\dd \lambda}\right)
    + \lambda^4\bar{\psi}(\lambda).
\end{equation*}

%%%%%%%%%%%%%%%%%%%%%%%%%%%%%%%%%%%%%%%%%%%%%%%%%%%%%%%%%%%%%%%%%%%%%%%%%%%%%%%%
\section{Acknowledgments}
This work is partially supported by the Scientific and Technical
Research Council of Turkey (TUBITAK), 2221-Fellowships for
Visiting Scientists and Scientists on Sabbatical Leave Programme.

%%%%%%%%%%%%%%%%%%%%%%%%%%%%%%%%%%%%%%%%%%%%%%%%%%%%%%%%%%%%%%%%%%%%%%%%%%%%%%%%
%

\end{document}